\documentclass[10pt]{iopart}
\usepackage{iopams}  
\usepackage{graphicx}

\def\thercsid{\relax}
\def\rcsid#1{\def\next##1#1{\def\thercsid{##1}}\next}
\rcsid$Id: cqg.tex,v 1.11 2004/05/04 18:07:57 jolien Exp $

\renewcommand{\today}{\number\day\space\ifcase\month\or
  January\or February\or March\or April\or May\or June\or
  July\or August\or September\or October\or November\or December\fi
  \space\number\year}

\begin{document}

\title[Upper limits on gravitational-wave signals]{%
Upper limits on gravitational-wave signals based on loudest events}
\author{Patrick R. Brady,
Jolien D. E. Creighton, and 
Alan G. Wiseman}
\address{Department of Physics, University of Wisconsin--Milwaukee,
  P.O. Box 413, Milwaukee, WI 53201}
\begin{abstract}
Searches for gravitational-wave bursts have often focused on the loudest
event(s) in searching for detections and in determining upper limits on
astrophysical populations.  Typical upper limits have been reported on
event rates and event amplitudes which can then be translated into
constraints on astrophysical populations.  We describe the mathematical
construction of such upper limits.
\end{abstract}


\submitto{\CQG}


\section{Introduction}

Data analysis pipelines designed to detect gravitational-wave bursts
yield event candidates.   While the current generation of
gravitational-wave detectors and the corresponding data analysis
pipelines are being commissioned, it is generally agreed that even a
very significant event would require substantial scrutiny before being
accepted as a detection of gravitational waves.  Nevertheless, data
from existing detectors can be searched for gravitational-wave bursts;
the results of the searches can be used to set upper limits on the
population of burst
sources~\cite{Nicholson:1996ys,Allen:1999yt,Tagoshi:2000bz,Abbott:2003pj,Abbott:2003pb}.
Furthermore, the choice of reporting an upper limit does not preclude the detection
of gravitational waves, if any are present with sizable amplitude
and/or rate.

For upper limits on the rate of burst events in the Universe,  a
traditional approach uses the number of event candidates that arise
from the pipeline as the observable.   Using this number,  and
information about the expected number of noise events arising from the
pipeline,  an upper limit on the rate of gravitational wave bursts can
be derived.   The dependence of the final limit on the discrete number
of observed event candidates can be a disadvantage in the cases where
the signals are weak and rare; small changes in the analysis pipeline
cuts can cause discrete changes in the reported rate limit.  Several
published upper limits
\cite{Nicholson:1996ys,Allen:1999yt,Abbott:2003pj} have used instead the
amplitude of the largest, or \emph{loudest}, event candidate to derive the reported upper
limit.  The loudest event method uses a continuous parameter,  related
to the significance of the event candidate,  to determine the upper
limit;   the upper limit then depends continuously on the observable.
A further advantage of the loudest event method is that the final
threshold level need not be determined in advance of the search.   In
searches for weak and rare events,   this means that the loudest event
method often provides the best rate limit.

In this paper, we describe the methodology used to obtain upper limits
by focusing on the most significant event candidate.  We derive
formulae for event rate upper limits, both Bayesian and frequentist, 
based on the loudest event candidate including the case when a
background rate can be reliably measured.  The frequentist version of
the loudest-event rate upper limit has been obtained previously by
Cousins~\cite{Cousins:1994gs} and related methods have been explored
in Ref.~\cite{Yellin:2002xd}.   When the probability of a background
event with amplitude greater than  the loudest event is assumed to be zero,
we show that the rate limit obtained is \emph{conservative} in the sense
that the bound is not violated in the presence of a background.  We
also show that setting an event rate limit based on the loudest event
candidate will often out-perform rate limits derived by counting
events above a pre-set amplitude threshold if the threshold choice is
not optimal (in the case that all candidate events are due to detector
noise).  We also describe how the loudest event can be used to set an upper
limit on the amplitude of the gravitational-wave strain observed from a
particular class of waveforms.

\section{Upper limits on event rate}

Typical searches for gravitational-wave bursts involve intricate
pipelines that process detector data and produce a list of candidate
events.  One of the parameters associated with each candidate event is
an estimate of the significance of the event candidate,  e.g., its
amplitude or confidence.  In a traditional event-counting approach to
rate estimation, a threshold on this significance parameter would be
used to distinguish between the events that are presumed to be of
astrophysical origin and those that are presumed to be due to detector
noise.

When setting an \emph{interpreted} upper limit on the rate,  the
detection efficiency of the search is also required.  The detection
efficiency $\epsilon(x)$ is a function of the amplitude $x$ and
represents the fraction of events from a (hypothetical)
astrophysical population that would produce event candidates with
amplitude greater than $x$ after processing through the analysis
pipeline.\footnote{Recall that the amplitudes are a description of the
output of the analysis pipeline, which is subject to detector noise as
well as the astrophysical signal, though it might be
related to some intrinsic amplitude of the gravitational waves
associated with the event.}   In some cases,  a population of signals
with fixed intrinsic strain-amplitude $h$ can be constructed and the
efficiency of detection for such a signal $\epsilon(x,h)$ can be
determined for a range of intrinsic signal strengths.  This allows
one to create a \emph{rate versus strain} plot often adopted as
the interpreted result for situations where a spatial distribution of
sources is not known~\cite{Abbott:2003pb}.

The detection efficiency is readily evaluated by Monte Carlo
simulation in which simulated signals are added to (usually real)
detector noise and analyzed through the pipeline.  In an
event-counting analysis, this efficiency is evaluated at a threshold
level $x^\ast$ and allows one to convert an uninterpreted bound on
event rate (which is an estimate of the rate of events that exceed
the threshold without regard to the origin of the events) into an
interpreted bound on the event rate, which requires knowledge of the
fraction of a hypothetical population that the detection pipeline
could have detected.

The detection efficiency has a slightly different role in the loudest
event method.  The detection efficiency
$\epsilon_{\mathrm{max}}=\epsilon(x_{\mathrm{max}})$,  evaluated at
the largest observed amplitude $x_{\mathrm{max}}$,  is the probability
that an astrophysical event would produce an amplitude greater than
$x_{\mathrm{max}}$ in the data analysis pipeline.   If
$\epsilon_{\mathrm{max}}$ is small,  then a large rate of events would
be needed to produce one of the rare event candidates with
significance greater than $x_{\mathrm{max}}$.   If
$\epsilon_{\mathrm{max}}$ is close to unity,  it is very likely that
an event would produce a more significant event candidate unless the
rate of events is sufficiently small.   This will be made
mathematically precise below.  It should be noted that the efficiency
is essential to the loudest event statistic---there is no way to
achieve an uninterpreted bound since without a hypothetical population
there is no way to know how rare the loudest event is.

In the absence of an event with sufficient significance to be
claimed a detection,  the loudest event method alleviates the
competing demands of setting a threshold that is sufficiently above
the level of detector noise while being not so large that the
detection efficiency becomes too low (and thus the upper limit
suffers).  Thus the loudest event method is appropriate when one
expects true events to be rare.
However, use of the loudest event is not incompatible with the goal of
making a detection. Indeed, one can simultaneously claim a detection
and use the loudest event to place an upper limit on the event rate.

\subsection{Rate upper limits without a background estimate}

If the population of astrophysical sources produces Poisson
distributed events with an intrinsic event rate $R$, then the probability that
all the event candidates have an amplitude less than some
value $x$ is determined by this intrinsic rate, the efficiency $\epsilon(x)$
evaluated at this value of the amplitude, and the observation time $T$ via
\begin{equation}
P(x|\mu) = \sum_{n=0}^{\infty} \frac{[1-\epsilon(x)]^n \mu^n e^{-\mu}}
{n!}   = e^{-\mu\epsilon(x)}
\end{equation}
where $\mu=RT$ is the Poisson mean.  
This formula can be used to determine a frequentist upper limit on the
rate $R$ by determining the largest amplitude $x_{\mathrm{max}}$
recorded by the data analysis pipeline,  and then solving
$1-p=P(x_{\mathrm{max}}|\mu_p)$ for $\mu_p=R_pT$ where $p$ is the
desired confidence level.   The result is
\begin{equation}
  R_p = -\frac{\ln(1-p)}{T\epsilon_{\mathrm{max}}}.
  \label{e:freq-lim-no-background}
\end{equation}
For example, a 90\%-confidence frequentist upper limit ($p=0.9$) is
\begin{equation}
\label{e:Frequentist90NoBackground}
  R_{90\%} = \frac{2.303}{T\epsilon_{\mathrm{max}}}.
\end{equation}
Thus,  an experiment will yield a value of $R_{90\%}$ that is less
than the true rate only 10\% of the time. 


A Bayesian upper limit is determined from the posterior probability
distribution $P(\mu < \mu_p| x_{\mathrm{max}})$ which can be computed
by a straightforward application of Bayes law.   The probability 
$p(x|\mu)\,dx$ of the loudest event being produced with
an amplitude between $x$ and $x+dx$ is first obtained by
taking the derivative of $P(x|\mu)$ with respect to $x$:
\begin{equation}
  p(x|\mu)=-\mu\epsilon'(x)e^{-\mu\epsilon(x)}
\end{equation}
where $\epsilon'(x)=d\epsilon(x)/dx$.  The posterior probability
distribution is then given by
\begin{equation}
P(\mu < \mu_p| x_{\mathrm{max}}) 
= {\mathcal{N}}^{-1} {\int_0^{\mu_p}d\mu\,p(\mu)p(x_{\mathrm{max}}|\mu)}%
\end{equation}
where ${\mathcal{N}} = {\int_0^\infty d\mu\, p(\mu)
p(x_{\mathrm{max}}|\mu)}$ is a normalization constant
and $p(\mu)$ is a prior distribution on the
event rate.  An upper limit on the rate at the $100p\%$ confidence level
is obtained by solving $P(\mu < \mu_p| x_{\mathrm{max}}) = p$ for
$\mu_p$.  For a uniform (improper) prior $p(\mu) =  \mathrm{constant}$, this
requires solving
\begin{equation}
  p = 1 - e^{-\mu_p\epsilon_{\mathrm{max}}}(1+\mu_p\epsilon_{\mathrm{max}})
\end{equation}
for $\mu_p$.  A $90\%$ Bayesian confidence level upper limit is then
\begin{equation}
  R_{90\%} = \frac{3.890}{T\epsilon_{\mathrm{max}}}.
\end{equation}
That is,  there is 90\% probability that the true rate is less than
$R_{90\%}$ given the observed value of $\epsilon_{\mathrm{max}}$ and
the prior assumption of a uniform rate distribution.

\subsection{Accounting for a background}

If the background distribution (i.e., the distribution of event
candidates that result from detector noise alone) is known, or can be
estimated,  then this information can be incorporated into the rate
upper limits derived above.  In
gravitational wave searches, this information is often obtained by coincidence
analyses performed on time-shifted data.  Let $P_0(x)$ be the probability that
all background events have an amplitude less than $x$ (as a function of $x$)
for the given search and observation time.  Since the
loudest event could have been either from the astrophysical foreground (which
is a Poisson process with mean $\mu$) or from the noise-produced background,
the probability that \emph{all} events lie below amplitude $x$ is
just the product
$P(x|\mu,\mathrm{B}) = P_0(x)e^{-\mu\epsilon(x)}$ where $\mu=RT$ and
$\mathrm{B}$ symbolically represents the inclusion of information about the
background.  Evaluated at the loudest event, this is
\begin{equation}
  P(x_{\mathrm{max}}|\mu,\mathrm{B})=P_0(x_{\mathrm{max}})e^{-\mu\epsilon_{\mathrm{max}}}
\end{equation}
and the rate upper limit at a frequentist confidence level $p$ is
\begin{equation}
  R_p = -\frac{\ln(1-p)-\ln P_0(x_{\mathrm{max}})}{T\epsilon_{\mathrm{max}}}.
  \label{e:freq-lim-with-background}
\end{equation}
Note that this rate limit will always be less than the rate limit in
Eq.~(\ref{e:Frequentist90NoBackground}) [or equal in the case that
$P_0(x_{\mathrm{max}})=1$].  This demonstrates that the no-background
assumption is conservative in the sense that the rate limit in
Eq.~(\ref{e:freq-lim-no-background}) will not be violated by failing to account
for the background.   Note,  however, that the rate in
Eq.~(\ref{e:freq-lim-with-background}) can become zero, or even negative,
when $P_0(x_{\mathrm{max}})\le1-p$.   This pathology is intrinsic to
the meaning of a frequentist upper limit;  a similar pathology occurs
in counting experiments if an unusually small number of events is
observed in a high background experiment.   A modified frequentist
statistic based on the loudest event might remove this pathology.  

The Bayesian upper limit, accounting for the background, is constructed as
before.  The probability $p(x|\mu,\mathrm{B})\,dx$ that the loudest event is produced
with amplitude between $x$ and $x+dx$ is obtained by taking the derivative of
$P(x|\mu,\mathrm{B})$ with respect to $x$:
\begin{equation}
  p(x|\mu,\mathrm{B})=[p_0(x)-\mu\epsilon'(x)P_0(x)]e^{-\mu\epsilon(x)}
\end{equation}
where $p_0(x)=dP_0(x)/dx$.   For a uniform prior, the $100\, p \%$
confidence upper limit is determined by solving 
\begin{equation}
  p = 1 - e^{\mu_p\epsilon_{\mathrm{max}}}
  (1+\xi\mu_p\epsilon_{\mathrm{max}})
\end{equation}
for $\mu_p$ where
\begin{equation}
  \xi = \left[1-\frac{\epsilon(x_{\mathrm{max}})}{\epsilon'(x_{\mathrm{max}})}
  \frac{p_0(x_{\mathrm{max}})}{P_0(x_{\mathrm{max}})}\right]^{-1}
  =\left[1-\left.\frac{d\ln P_0}{d\ln\epsilon}\right|_{\epsilon_{\mathrm{max}}}
  \right]^{-1}
\end{equation}
is the correction accounting for the background.
[The second expression for $\xi$ can be obtained by expressing the background
probability $P_0$ as a function of $\epsilon$ rather than $x$, which requires
inverting the function $\epsilon(x)$].  Notice that $0\leq
\xi \leq 1$.     When $\xi=1$,  i.e.  $p_0(x_{\mathrm{max}})=0$ and
$P_0(x_{\mathrm{max}})=1$, the Bayesian upper limit in the case of no
background is recovered.    Thus the assumption underlying the
no-background limit is that there is no chance that a background event
could exceed $x_{\mathrm{max}}$.  Otherwise, the Bayesian upper limit
will always be smaller than the limit when no background was assumed;
i.e., the no-background limit is again a conservative
limit.  In the extreme case of the loudest event almost certainly
being due to the background ($\xi=0$), the rate upper limit becomes
\begin{equation}
  R_{90\%} = \frac{2.303}{T\epsilon_{\mathrm{max}}}.
\end{equation}

In the special case of a Poisson-distributed background with mean $\mu_0(x)$
number of events with amplitudes greater than $x$, the correction $\xi$
is~\cite{finn}
\begin{equation}
  \xi = \left[1-\frac{\epsilon(x_{\mathrm{max}})}{\epsilon'(x_{\mathrm{max}})}
  \mu'_0(x_{\mathrm{max}})\right]^{-1}
\end{equation}
where $\mu'_0(x)=d\mu_0(x)/dx$.

\subsection{Rate-versus-strain plots}

In some circumstances it is not possible to construct a definitive model
population, yet an interpreted bound on rate is possible by setting a rate
as a function of some model parameter.  For example, in
Ref.~\cite{Abbott:2003pb}, the detection efficiency is computed for 
a particular waveform as a function of the intrinsic waveform strain $h$,
and this is used to construct a rate-versus-strain plot.  The loudest event
rate limits described so far can be immediately generalized to this case
by allowing the efficiency to be a function of the parameter $h$,
$\epsilon=\epsilon(x,h)$.  Note that $x$ represents the signal strength
(or significance) as measured by the pipeline while $h$ is a measure
of the intrinsic signal strength as defined by the population.  (In fact,
$h$ could represent any set of parameters characterizing the population.)
The frequentist rate-versus-strain plot (with background) is given simply by
Eq.~(\ref{e:freq-lim-with-background}) where
$\epsilon_{\mathrm{max}}=\epsilon_{\mathrm{max}}(h)$ is treated as a function
of the population parameter $h$.

\subsection{Advantage of the rate limit based on the loudest event}

\begin{figure}
\begin{center}
\includegraphics[width=\linewidth]{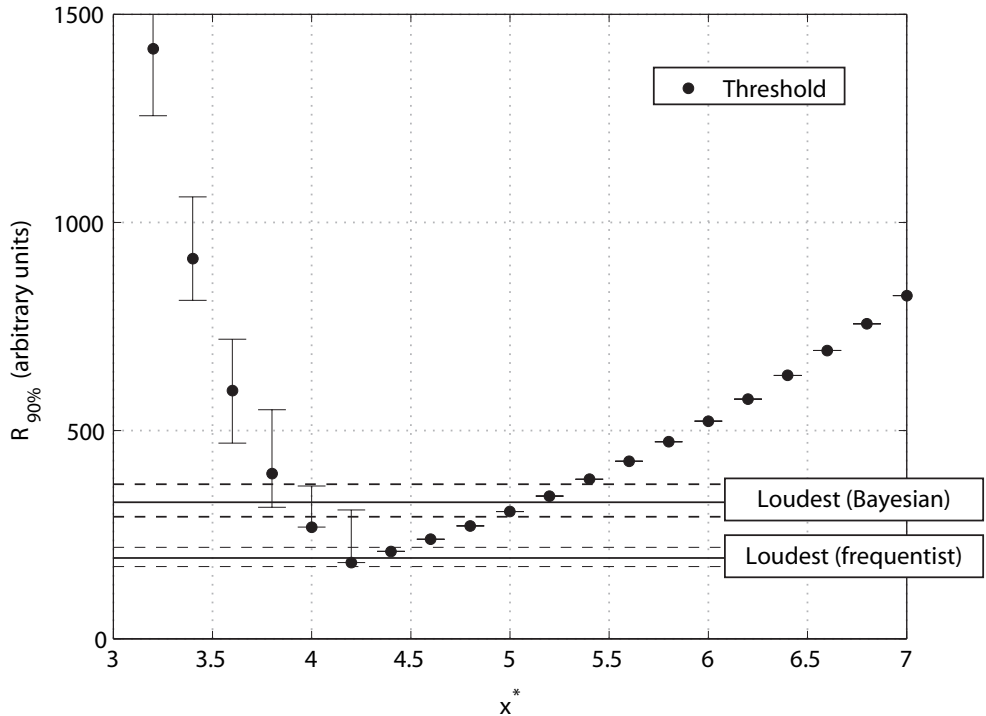}
\end{center}
\caption{\label{f:threshold}%
The median value and interquartile ranges of the frequentist 90\%
confidence upper limit on the rate of a Poisson process as a function
of amplitude threshold $x^\ast$ based on counting the number of events
and ignoring the noise background.  The results are based on a  Monte-Carlo
simulation involving 10000 instances of 10000 noise events drawn from a
probability distribution of the form $P_0(x) \sim e^{-x^2/2}$ assuming
an efficiency which varies as $\epsilon \sim x^{-3}$.   Also shown on
the figure are the median values and interquartile ranges for the frequentist 
and Bayesian 90\% confidence upper limits on the rate using the
loudest event.   These results are shown as lines for ease of
comparison since there is no threshold dependence.  Notice the narrowness
in the choice of threshold $x^\ast$ that would give rise to an upper limit
that is competative with the limit provided by a loudest event statistic.}
\end{figure}

The choice of the threshold required for an event counting approach to
constructing an upper limit necessitates a balance between the competing goals
of setting the threshold high enough that one is not dominated by a large
number of false alarms (which would result in an artificially
high upper limit) and setting the threshold low enough that the detection
efficiency is not compromised (which would again result in an artificially high
upper limit).  In the case that there is no true signal present, the loudest
event statistic represents, in some sense, tuning the threshold to the nearly
optimal level: right at the loudest event.  To illustrate this, consider the
case in which the noise consists of a large number $N$ background events where
the probability that any single event has an amplitude less than $x$ is
$e^{-x^2/2}$ so that $P_0(x)\sim e^{-x^2/2}$. (This would be the expected
behavior for the problem of detecting binary inspirals in Gaussian noise.)
Assume that the detection efficiency is $\epsilon\sim x^{-3}$ for large
amplitude (as would be expected for gravitational wave sources distributed
homogeneously in space).  A Monte Carlo realization of 10000 instances of noise
with each instance consisting of 10000 noise events,
shows the upper limit that would be obtained for various choices of threshold
via an event counting method, see Fig.~\ref{f:threshold}.  Also shown on this
plot is the upper limit that would be obtained using a conservative loudest
event method (i.e., with no background accounting) for both the Bayesian and
frequentist statistics.  (These are shown as horizontal lines in the plot
although these statistics are \emph{not} functions of a threshold.)  This plot
shows that the upper limit obtained by a counting method with a threshold
depends rather sensitively on the threshold level, and that the loudest event
methods perform nearly as well as possible unless a nearly perfect threshold
level can be achieved.

\section{Upper limits on event strength}

Another type of gravitational wave search does not seek to bound a rate
on the number of events occurring during the observational period but rather
to set a limit on the strength of any single event that may be present 
during that observation time.  A significantly strong candidate may be
described as a detection independently of such an upper limit.  Such
upper limits have been sought in past gravitational-wave
searches~\cite{Nicholson:1996ys,Abbott:2003yq}.  The crucial
assumption in such an approach is that a single event is present, as would
be the case for a triggered search (e.g., a search for a gravitational wave
burst associated with an observed supernova) or for a search for a known
pulsar.  
The central quantity is then the probability that all triggers have
amplitude less than the observed maximum $x_{\mathrm{max}}$ given the
presence of a signal with intrinsic strain $h$,  i.e.,  $P(x < 
x_{\mathrm{max}}| h)$ [this is what was previously denoted
$1-\epsilon_{\mathrm{max}}(h)$].    This probability distribution can be
estimated using injections into off-source times (or frequencies),
then the $100p\%$ frequentist upper limit on the signal strain
is determined by solving
\begin{equation}
(1 - p) = P(x < x_{\mathrm{max}}| h_p)
\label{e:strain-limit}
\end{equation}
for $h_p$.    Notice that there are some problems with this limit.
Suppose,  for example,  that $x_{\mathrm{max}}$ is unusually small
in the sense that the $P(x < x_{\mathrm{max}}| h=0)$ is
small.   Then the value of $h_p$ might be abnormally low.
[This happens when the false alarm probability
$1-P(x<x_{\mathrm{max}}|h=0)$ is close to the required confidence, $p$,
for the upper limit.]
Once again,  this is an artifact of the frequentist approach to the upper
limit and could be avoided by a Bayesian analysis.

\section{Summary}

We have described several upper-limit statistics based on the loudest
observed event.  These statistics can be used to bound the rate of events
coming from a known source population, to construct a rate-versus-strain
exclusion curve, or to bound the amplitude of an event if at most one event
is known to be present.  The loudest event methods have advantages over
traditional methods that involve counting events that exceed a threshold in
that the loudest event methods: (i) alleviate the (often difficult) task of
threshold choice, (ii) use a continuous parameter (the amplitude of the loudest
event) rather than a discrete parameter (the number of events above threshold),
and (iii) will usually out-perform the event counting method (meaning that the
upper limit will be more constraining at a given confidence level).

A number of alternatives to the loudest event method for use in gravitational
wave data analysis remain to be explored.  For example, a
``next-loudest event'' statistic may be more robust against a potentially
damaging single noise glitch.  An interesting approach has been developed by
Yellin~\cite{Yellin:2002xd}, which establishes a limit based on the largest
gap in the efficiency between events.

\ack

We would like to thank Peter Shawhan for several useful comments on
this work, and for bringing our attention to
Refs.~\cite{Cousins:1994gs} and~\cite{Yellin:2002xd}.  This work has
been supported by the National Science Foundation Grant PHY-0200852.
Patrick Brady is also grateful to the Alfred P Sloan Foundation and
the Research Corporation Cottrell Scholars Program for support.

\end{document}